\documentclass[twocolumn,twoside,slac_two]{revtex4}
\usepackage{graphicx}
\usepackage{fancyhdr}
\begin{document}

\title{On the Positron Fraction and Cosmic-Ray Propagation Models}

\author{B. Burch}
\affiliation{Washington University in St. Louis, St. Louis, MO 63130, USA}
\author{R. Cowsik}
\affiliation{Washington University in St. Louis, St. Louis, MO 63130, USA}

\begin{abstract}
The positron fraction observed by PAMELA and other experiments up to $\sim100$ GeV is analyzed in terms of models of cosmic-ray propagation. It is shown that generically we expect the positron fraction to reach $\sim0.6$ at energies of several TeV, and its energy dependence bears an intimate but subtle connection with that of the boron to carbon ratio in cosmic rays. The observed positron fraction can be fitted in a model that assumes a significant fraction of the boron below $\sim10$ GeV is generated through spallation of cosmic-ray nuclei in a cocoonlike region surrounding the sources, and the positrons of energy higher than a few GeV are almost exclusively generated through cosmic-ray interactions in the general interstellar medium. Such a model is consistent with the bounds on cosmic-ray anisotropies and other observations.
\end{abstract}
\maketitle

\thispagestyle{fancy}

\def\bd{\begin{displaymath}}
\def\be{\begin{equation}}
\def\ed{\end{displaymath}}
\def\ee{\end{equation}}

\section{Introduction}
\begin{figure}
\includegraphics[width=7cm]{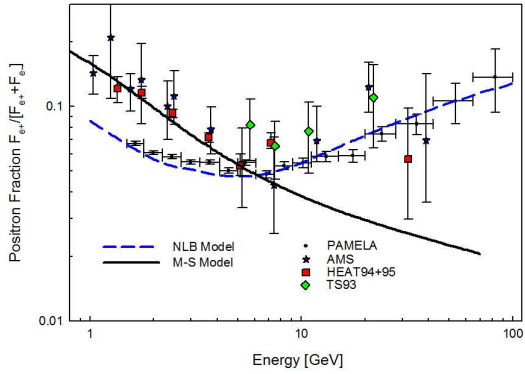}
\caption{The positron fraction measured by PAMELA and earlier measurements are shown. Gradient drifts in solar modulation may account for some of the difference in the data at $E<10$ GeV \cite{Adriani}. The prediction of the positron fraction expected in the M-S model is shown as a solid line and in the NLB model as a dashed line.}\label{PAM}
\end{figure}

The recent observation of the positron fraction in cosmic rays by PAMELA \cite{Adriani} has created much excitement because of its possible connection with the annihilation or decay of dark matter in the Galaxy or with a variety of astrophysical processes (see \cite{PRD} for references to these discussions). 
These suggestions were prompted by the recognition that the energy dependence of the positron fraction cannot be fitted by the comprehensive propagation model (solid line in Fig. \ref{PAM}) developed by Moskalenko and Strong (M-S) \cite{Moskalenko,Strong}. In Fig. \ref{PAM}, we show the PAMELA observations of the ratio, $R$, of the positron flux to that of the total electronic component in cosmic rays along with earlier observations \cite{AMS,HEAT,TS93}. The PAMELA measurements have been called anomalous as they do not conform to the predictions of the M-S model. Accordingly, new models of cosmic-ray propagation have been discussed (see references in \cite{PRD}). 


General arguments based on cosmic-ray propagation models indicate that the positron fraction should increase at high energies and asymptotically reach a value of $\sim0.6$ at the highest energies. We note that $\gamma$-ray astronomy has shown that cosmic rays generated in the sources suffer nuclear interactions in the proximity of the sources \cite{Malkov,Zhang,Fang,Funk,Acciari,Abdo4,Abdo2,Abdo3}. This has a strong bearing on the models of cosmic-ray propagation in that if a fraction of the $B/C$ ratio observed in cosmic rays, especially at energies below $\sim10$ GeV, is generated in a dense cocoonlike region surrounding the sources, then the contribution from spallation in the general interstellar medium would have a flat or a weak dependence on energy. Such a model \cite{Cowsik73,Cowsik75,Cowsik09} is shown to fit the PAMELA observations and to be consistent with the high degree of isotropy observed in cosmic rays at high energies \cite{Strong,Antoni,Abbasi,Tibet}.


\section{Positron Fraction at High Energies}

The asymptotic value of the positron fraction is estimated by noting that cosmic rays observed near the Earth are accelerated in a set of discrete sources distributed over the Galaxy \cite{Cowsik79,Nishimura}, which accelerate mostly electrons rather than positrons, as the Galaxy is made up of matter rather than antimatter. During the diffusive transport, the electronic component suffers loss of energy due to synchrotron radiation and inverse-Compton scattering on the microwave background and other photons. As this loss increases quadratically with energy as $bE^2$, the spectrum of the electronic component is sharply cut off  at high energies. Solutions to the diffusion equation \cite{Cowsik09,Cowsik79}, which include the energy losses by electrons, yield a spectrum that cuts off as
\be F_e(E,r_n)\sim exp \Bigg(-\frac{br_n^2EE_x}{4\kappa(E_x-E)}\Bigg).\ee
Here, $r_n$ is the distance to the nearest source, $E_x$ is the maximum energy up to which the sources accelerate electrons, the diffusion constant $\kappa\approx10^{28}$ cm$^2$s$^{-1}$, and $b\approx1.6\times10^{-3}$ GeV$^{-1}$Myr$^{-1}$. Thus even for a very large value of $E_x$, the directly accelerated electron spectrum is cut off at $E_b\approx4\kappa/(br_n^2)\approx100$ GeV/(r$_n$/kpc)$^2$. The cutoff in the spectrum at $\sim$ 1 TeV observed by the HESS instrument \cite{HESS} indicates the presence of cosmic ray sources within $\sim$200 pc of the solar system. If this is taken to be the typical spacing  between the sources in our Galaxy, then we expect about $10^4$ sources in this disk within a radius of $\sim$15 kpc \cite{Cowsik79}; accordingly, each of these sources need only to generate a very small fraction of the cosmic ray luminosity of the Galaxy, on the average.

We do not expect the secondary electrons and positrons to exhibit such a cutoff because, unlike the discrete sources of primary electrons, the source function for the secondary component extends from the nearest proximity to the solar neighborhood to far-off distances. The secondary positrons and electrons are generated through the $\pi^\pm\rightarrow\mu^\pm\rightarrow e^\pm$ decay chain, the pions being produced in high-energy interactions of cosmic rays with the matter in interstellar space, both of which are distributed rather smoothly, without large overall gradients. Accordingly, the effects of the energy loss are less severe and the index of the secondary electron spectrum at low energies is the same as the source spectrum, which is the same as that of the nucleon spectrum \cite{Cowsik66,Protheroe,Stephens}. At high energies, the secondary spectra of positrons and electrons steepens by one additional power.
\begin{eqnarray}
F_s(E)&\sim&E^{-\beta}~~~~~~~~~for~~~E\ll E_c\nonumber\\
F_s(E)&\sim&E^{-(\beta+1)}~~~~for~~~E\gg E_c
\end{eqnarray}
where $\beta=2.65$ and $E_c\sim100-200$ GeV.

This spectrum of the secondary electronic component will progressively dominate over that generated by the discrete sources. This implies that at very high energies, the positron fraction simply corresponds to that in the production process in the high-energy collisions of cosmic rays. The fact that the $p/n$ fraction in primary cosmic rays is greater than unity favors the production of $e^+$ over $e^-$, reflecting the slightly greater production of $\pi^+$ compared with $\pi^-$. Whereas the theoretical calculations \cite{Protheroe,Stephens} yield $F_{s+}/F_{s-}\approx1.5-2$, the direct observations of $\mu^+/\mu^-$ produced by cosmic rays in the Earth's atmosphere yields $F_{s+}/F_{s-}\approx1.3$ \cite{Hayakawa}. Then, for $E\gtrsim1$ TeV
\be R_s(E)\rightarrow\frac{F_{s+}(E)}{F_{s+}(E)+F_{s-}(E)}\approx0.6.\ee
We may expect that such a large value of $R$ will be reached at $E>1$ TeV, say beyond several TeV.

\section{Energy Dependence at Moderate Energies}

\subsection{Residence Time of Cosmic Rays}

\begin{figure}
 \includegraphics[width=7cm]{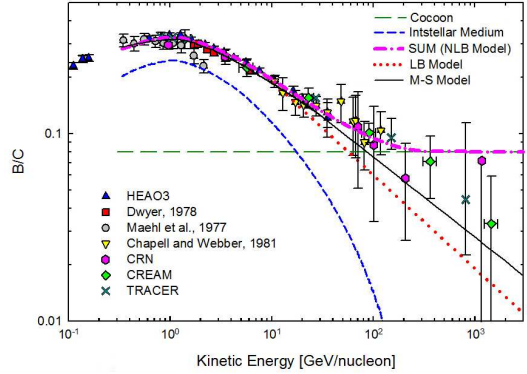}
  \caption{The observed $B/C$ ratio \cite{HEAO3,Dwyer,Maehl,Chapell,Tracer,CRN,Cream} with the predictions from the M-S and NLB models are shown.}\label{BCSC}
\end{figure}

There are two classes of models for cosmic-ray propagation with which to explain the measurements of the primary and secondary nuclei in cosmic rays as show in Fig. \ref{BCSC}. In the M-S model, the secondary production is distributed throughout the Galaxy, and the observed decrease with energy of the ratio of secondary to primary nuclei is explained by an effective residence time of cosmic rays in the Galaxy decreasing with energy \cite{Moskalenko,Strong,Jones}. This decrease may be parameterized beyond a few GeV/n by
\be\tau_{L}(E)\sim\tau_{L0}(E_0+E)^{-\Delta}\approx\tau_{L0}(E_0+E)^{-0.5} \ee
where $E_0\approx4$ GeV/nucleon, and we have indicated $\Delta\approx0.5$ to reflect the full range $0.33\leq\Delta\leq0.7$ of the M-S models currently under discussion in literature. The value of $\tau_{L0}\approx0.4$ is in units of $T_0$, and $E$ and $E_0$ are expressed in GeV. Models of this class, which may be approximated by a leaky-box (LB) model \cite{Cowsik66,Cowsik67}, produce a nuclear secondary to primary ratio such as that given by the dotted lines in Fig. \ref{BCSC}. Note here that the LB model approximates the predictions of the M-S model also shown in Fig. \ref{BCSC}. The second class of models takes explicit account of significant secondary production in dense regions in the vicinity of the primary cosmic-ray sources. Such a model may be realized as a nested leaky-box (NLB) \cite{Cowsik73,Cowsik75}.

\subsection{Including Spallation in the Source Regions}

In the NLB model, it is assumed that subsequent to the acceleration,  the cosmic rays spend some time in a cocoon-like region surrounding the sources, interacting with matter and generating some of the secondaries, mainly at low energies. Such interactions will also generate gamma rays through the $\pi^\circ\rightarrow2\gamma$ decay and could be observed by space-borne gamma-ray telescopes like FERMI \cite{Tibaldo,Abdo1}. Since, according to the arguments summarized in Section 2, the average luminosity of a cosmic-ray source is rather low, their gamma-ray emission will be detected only in some favorable cases. The effective residence time in the cocoon, $\tau_c(E)$, is energy dependent, with the higher energy particles leaking away more rapidly from the cocoon. After they leak out of the cocoon into the interstellar medium, the cosmic rays at all energies up to several hundred TeV reside for an effective time $\tau_G$ before they escape from the Galaxy. In the NLB model, the observed energy dependence of the nuclear secondary to primary ratio is fit with an energy-dependent leakage time $\tau_c(E)$ out of the cocoon and with a leakage time $\tau_G$ out of the Galaxy that is independent, or nearly independent, of energy up to $\sim$1 PeV. These two contributions are depicted by the dashed lines in Fig. \ref{BCSC}, and their sum is shown as a chain-dotted line. This shows that the residence time for cosmic rays inside the cocoon has a progressively steeper dependence on energy, and $\tau_{c}(E)$ and $\tau_G$ may conveniently be parameterized as
\begin{eqnarray}
\tau_c(E)&\sim&\tau_{C0}E^{\epsilon-\delta log E},\quad\tau_G\sim~constant\nonumber\\
\tau_{N}&=&\tau_c(E)+\tau_G.
\end{eqnarray}
Here, the lifetimes $\tau_C$, $\tau_G$, and $\tau_N$ are in $T_0$ units and take on values $\tau_{c0}\approx0.24$ and $\tau_G\approx0.08$ when $E$ is expressed in GeV, with the parameters $\epsilon=-0.01$ and $\delta=0.13$. 
The cocoon should have a high density so that adequate spallation might take place in the short amount of time that the cosmic rays spend around their sources. Circumstellar envelopes, dark clouds, molecular clouds, and giant molecular complexes are some of the candidates that may serve as cocoons. These have widely ranging densities, from $\sim10^7$ cm$^{-3}$ down to $\sim10^2$ cm$^{-3}$ \cite{Allen}, and the cosmic rays need to spend anywhere from 10 yr to $10^5$ yr in these regions to generate the requisite $B/C$ ratio at $\sim1$ GeV. Since the dimensions of these regions are inversely correlated with their densities, such residence time in the cocoon may be generated with diffusion constants in the range of $10^{26}-10^{28}$ cm$^2$ s$^{-1}$.

\begin{figure}
\includegraphics[width=7cm]{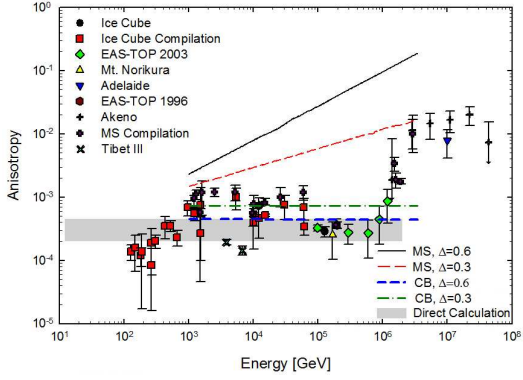}
\caption{Measurements of the cosmic-ray anisotropy from various compilations \cite{Strong,Antoni,Abbasi,Tibet}. Also plotted are the predictions from models in Moskalenko and Strong (MS) \cite{Strong} and the results from Eq. 6 (CB). The gray region shows the predicted anisotropy from Eq. 7.}\label{Aniso}
\end{figure}

Both LB and NLB models can provide adequate fits to the nuclear secondary to primary ratios observed to date, even though the difference between them becomes progressively larger at higher energies. Whereas the LB models require an effective galactic residence time, $\tau_L(E)$, progressively decreasing with energy, the NLB models fit the data on cosmic-ray nuclei with a constant residence time $\tau_G$ at high energies. Accordingly, LB models predict cosmic-ray anisotropies that increase with increasing energy, in conflict with the observations \cite{Strong,Antoni,Abbasi}. In contrast, NLB models predict constant anisotropies up to several hundred TeV, consistent with the observations as shown in Fig. \ref{Aniso}. To be specific, the expected anisotropies, $\delta(E)$, are inversely proportional to the effective residence time of cosmic rays in the Galaxy. Accordingly, the anisotropy in the NLB model $\delta_{NLB}(E)$ is given by
\begin{eqnarray} \delta_{NLB}(E)&=&\frac{\tau_{LB}(E)}{\tau_G}\delta_{M-S}(E)\nonumber\\ &\approx&\Bigg(\frac{100~GeV}{E}\Bigg)^{\Delta}\delta_{M-S}(\Delta,E).
\end{eqnarray}
Here, $\delta_{M-S}(\Delta,E)$ refers to the anisotropy in the M-S model calculated for the two values $\Delta\approx0.3$ and $\Delta\approx0.6$ \cite{Strong}, and 100 GeV refers to the energy at which $\tau_L(E)$ and $\tau_G$ intersect in Fig. \ref{BCSC}. When their estimates are rescaled for the NLB model, according to Eq. 6, the expected levels of anisotropy become consistent with the observational limits \cite{Strong,Antoni,Abbasi,Tibet}.

We can also directly estimate the anisotropy parameter $\delta_{NLB}$ using the standard formula in cosmic-ray literature \cite{Strong}
\be \delta_{NLB}=\frac{3\kappa\nabla\rho}{c\rho}\approx\frac{3\kappa}{h_0c}\approx3\times10^{-4}\ee
where $h_0\approx1$ kpc$\approx3\times10^{21}$ cm is the scale height of the distribution of the cosmic rays above the Galactic plane and $\kappa\approx10^{28}$ cm s$^{-2}$ is the diffusion constant of cosmic rays in the interstellar medium. This anisotropy is shown in Fig. \ref{Aniso} with an uncertainty of $\sim200\%$ as a gray band. Note this estimate matches the values of anisotropy scaled down from the M-S calculations using Eq. 6. Below $\sim1$ TeV, the magnetic fields in the solar system, anchored at the Sun, prevent dipole anisotropies from being observed. Above $\sim1$ PeV, as we approach the knee in the cosmic-ray spectrum, the particles escape with increasing rapidity from the galactic volume, causing the anisotropy to increase. Keeping these factors in mind, we note that the anisotropy levels expected in the NLB model is consistent with the observations.

Another difference between the two models is that they require different input spectra to be generated by the sources. To see this, let $s_n(E)$ represent the spectrum of nuclei accelerated by the source by written as
\be s_n(E)=s_{n0}E^{-\alpha}.\ee
Since in the LB model these nuclei have a lifetime $\tau_L(E)$ and an interaction lifetime $(cn_H\sigma_{int})^{-1}$, the spectrum of cosmic-ray nuclei in the interstellar space becomes
\begin{eqnarray}
F_n(E)&=&s_n(E)\frac{\tau_L(E)}{1+c\sigma_{int}n_H\tau_L(E)}\\
\approx s_n(E)\tau_L(E)&=&s_{n}\tau_{L0}(E_0+E)^{-\Delta}=s_{n0}\tau_{L0}E^{-\alpha-\Delta}\nonumber
\end{eqnarray}
for $E\gg E_0$.

Here $(c\sigma_{int}n_H)^{-1}$ is the effective mean free path for the loss of cosmic rays at a particular energy through nuclear interactions. In order to match the observed spectrum of cosmic rays with $F_n(E)=F_0E^{-\beta}$ in the LB model, we need to set
$\alpha+\Delta=\beta\approx2.67$.
Thus
\be s_n(E)=s_{n0}E^{-\beta+\Delta}\sim E^{-2.2}.\ee

The calculation of the source spectrum in the NLB model, $s_{nc}(E)=s_{c0}E^{-\xi}$, is a two-step process. The spectral density inside the cocoons is the product of $s_{nc}(E)$ and the leakage lifetime inside the cocoon:
\be F_c(E)=s_{nc}(E)\tau_c(E)=s_{nc}(E)\tau_{c0}E^{\epsilon-\delta\log E}.\ee
These leak out into the interstellar space at a rate inversely proportional to the leakage lifetime from the cocoon so that
\be F_n(E)=F_c(E)\frac{1}{\tau_c(E)}=s_{nc}(E)=s_{c0}E^{-\xi}.\ee
Thus to match with the observed spectrum of cosmic rays with $F_n(E)\sim E^{-\beta}$ we need $\xi=\beta$.
This means that the observed cosmic rays have spectra identical to that accelerated by the sources, especially at high energies where losses due to ionization and nuclear interactions are small in the source regions.

\subsection{Derivation of the Positron Fraction}

In assessing the positron fraction in the NLB models, we note that the secondary nuclei, such as $B$, which are generated by the spallation of primary nuclei like $C$, have the same energy per nucleon as their progenitors. By contrast, positrons carry away, on the average, only about $5\%$ of the energy per nucleon of their nuclear progenitors \cite{Moskalenko,Protheroe,Stephens}. This implies that even though a significant amount of $B$ is generated through spallation within the cocoon, very little production of positrons at energies beyond $5$ GeV occurs there. This is because the progenitors of the positrons with $E\geq$5 GeV should have energies beyond about $100$ GeV per nucleon and would rapidly leak out of the cocoon before they suffer significant nuclear interactions (see dashed line in Fig. \ref{BCSC}). Thus, in the M-S model with energy-dependent path-length distributions, and in the NLB models, we expect the \emph{source function} for the positrons to be the same -- it is simply proportional to the product of the observed spectrum of the cosmic-ray nuclei and the density of the interstellar medium and has the spectral form $s_0E^{-\beta}$. Below $\sim$100 GeV, where the radiative energy losses are not significant, the observed positron fluxes would be the product of this source function and the residence time of cosmic rays in the Galaxy, $\tau_L(E)$ or $\tau_G$, as relevant to the model under consideration.

The calculation of the positron ratio in the two classes of models is straightforward when we note that its source spectrum $S_{+G}(E)$ in the interstellar medium is generated through nuclear interactions \cite{Protheroe} and has a nearly identical spectrum to that of the parent nuclei, $F_0E^{-\beta}$, except that it is shifted down in energy by a factor $\eta\approx0.05$ and multiplied by the rate of nuclear interactions
\be S_{+G}(E)\approx\sigma_{in} n_H F_0 \eta^{\beta-1}E^{-\beta}.\ee
Here, $\sigma_{in}$ is the inclusive cross section for the production of $\pi^+$, which carries off a fraction $\eta$ of the energy per nucleon of the primary cosmic-ray nucleus. The factor $\eta^{\beta-1}$ in Eq. 13 accounts for the shift in the energy and the change in the energy bandwidth when transforming from the spectrum of the primary nuclei to that of the positrons. The source function is the same for both the M-S and NLB models. In the NLB model, there is an additional small contribution $S_{+c}$ due to positron generation from nuclear interactions in the cocoon. Taking the expression for the spectral density for the nuclei in the cocoon from Eq. 11 shows
\begin{eqnarray}
S_{+c}(E)&\approx&c\sigma_{in}n_{Hc}s_{nc}\eta^{-\epsilon+\beta-1+\delta log(E/\eta)}\nonumber\\
&&\times E^{\epsilon-\beta-\delta log(E/\eta)}.
\end{eqnarray}
However, this contribution is entirely negligible beyond a few GeV. Therefore the steady state spectra $F_{+LB}(E)$ and $F_{+NLB}(E)$ are essentially given by the product of the source function and the effective lifetime of the positrons in the Galaxy. At energies below $\sim100$ GeV the radiative losses are small and the effective lifetimes in the two models are essentially given by the leakage lifetime $\tau_{L}(E)$ or $\tau_{NLB}\approx\tau_G$, respectively. Thus the positron spectra in the two models are given by
\begin{eqnarray}
F_{+LB}&=&S_{+G}(E)\tau_{L}(E)\nonumber\\
&\sim& E^{-\beta}(E_0+E)^{-\Delta}\sim E^{-(\beta+\Delta)};\\
F_{+NLB}(E)&=&S_{+G}(E)\tau_G\sim E^{-\beta}.
\end{eqnarray}


In order to estimate the positron fraction in the two models, we divide the positron fluxes $F_{+LB}$ and $F_{+NLB}$ by the spectral intensities of the total electronic component in cosmic rays. A recent compilation of the observations of the total electronic component can be found along with a smooth fit to the data that includes a slight enhancement in the intensities below $\sim 1$ GeV, which corrects for the effects of modulation by the solar wind can be found in Cowsik and Burch \cite{Cowsik09}. It is straightforward to take ratios of these spectra and compare the theoretically expected positron fraction in the two models with the observations shown in Fig. \ref{PAM}. At high energies, the spectra of positrons being essentially power laws in both the LB and NLB models, the shape of the positron fraction is controlled by the spectrum of the total electronic component. Thus, at high energies, we have
\begin{eqnarray}
R_{NLB}&\sim&S_+(E)\tau_G/F_{\pm}(E)\nonumber\\
&\sim& E^{-2.65}/E^{-3.1}\sim  E^{0.45}\\
R_{M-S}&\sim&S_+(E)\tau_L/F_{\pm}(E)\nonumber\\
&\sim& E^{-2.65}E^{-\Delta}/E^{-3.1}\sim E^{-0.2}
\end{eqnarray}
We see in Fig. \ref{PAM} that the NLB model shows the positron fraction increasing with energy at high energies and the M-S model shows a declining positron fraction at high energies.

\section{Conclusions}
We see that the nested leaky-box model provides a satisfactory fit to the PAMELA observations. This analysis obviates the need for exotic sources of positrons, suggested by comparison between the PAMELA data and the M-S propagation model, and shows that the data may be accounted for by NLB propagation models. Since NLB models also relieve the anisotropy problem encountered in the LB/M-S class of models and qualitatively accommodate the observations of $\pi^0$ gamma rays from regions near cosmic-ray sources, we conclude that the rising positron fraction observed by PAMELA is the natural result of cosmic-ray interactions in the interstellar medium.

\bigskip
\begin{acknowledgments}
We would like to thank M. H. Israel for his insightful suggestions.
\end{acknowledgments}

\end{document}